\documentclass[
	a4paper, 
	10pt, 
	unnumberedsections, 
	twoside, 
]{LTJournalArticle}

\runninghead{} 

\usepackage{amsmath}

\title{Enabling Syscall Interception on RISC-V} 

\author{%
	Petar Andrić,
    Aaron Call,
    Ramon Nou
}

\date{\footnotesize Barcelona Supercomputing Center}

\renewcommand{\maketitlehookd}{%
	\begin{abstract}
        The European Union's technological sovereignty strategy centers around the RISC-V Instruction Set Architecture, with the European Processor Initiative leading efforts to build production-ready processors. Focusing on realizing a functional RISC-V ecosystem, the BZL initiative (www.bzl.es) is making an effort to create a software stack along with the hardware. In this work, we detail the efforts made in porting a widely used syscall interception library, mainly used on AdHocFS (i.e., DAOS, GekkoFS), to RISC-V and how we overcame some of the limitations encountered. 
    \end{abstract}
}

\begin{document}

\maketitle 

\section{Introduction and motivation}
The RISC-V Instruction Set Architecture (ISA) is at the core of the European Union's technological sovereignty plans. A key initiative supporting this goal is the European Processor Initiative (EPI)~\cite{epi}, which aims to develop processors based on the open-source RISC-V ISA for mass production. These processors are intended for various applications, including cloud computing and data centers.

One of the storage innovations of the last years has been the ephemeral storage system~\cite{Brinkmann2020}. Those storage file systems use local resources to create a fast distributed file system that can be used to cope with the limitations of the general parallel file system. Those big storage systems suffer from interference~\cite{10820644} from the high number of jobs, users, and workloads that use them. Having a virtual file system capable of overcoming these limitations for the running job is a way to increase the storage efficiency of new systems and applications.

However, having an alternative file system on an HPC computer is not an easy task. Normally, we need permissions that are higher than those of the normal user, and system administrators limit such opportunities. To solve that, ephemeral file systems tend to use interception systems, \textit{libc} interception being the most standard. \textit{Libc} interception has some limitations, like reduced compatibility as we need to implement all the \textit{libc} calls using a file system. Finally, if the application or other libraries bypass \textit{libc}, adding extra calls becomes very difficult and hard to debug. To circumvent this, Intel\footnote{\label{Intel_syscall} \url{https://github.com/pmem/syscall_intercept}} developed a syscall intercept library that dynamically patches the binaries and modifies the system calls. This opens the opportunity to call pre- and post-operations and reimplement them at the user level, resulting in the creation of high-performance file systems on top.

In this poster, we explain our work to enable the syscall intercept library on top of RISC-V hardware\footnote{\scriptsize \url{https://github.com/GekkoFS/syscall_intercept/tree/riscv}}, following the previous efforts to port it to ARM (by RIKEN\textsuperscript{\ref{Intel_syscall}}) and to port it to PowerPC by our institution\footnote{\scriptsize \url{https://github.com/GekkoFS/syscall_intercept/tree/powerpc}}.

\section{Implementation}
The implementation on x86, PowerPC, and ARM architectures differs significantly from the one on RISC-V. Therefore, the obstacles, the solution, and the trade-offs are described in this section.

\label{subsec:obstacles}
\subsection{Obstacles}
\begin{enumerate}
    \item The relative jump instruction (\texttt{jal}) has a reach of $\pm$1\,MiB, which is insufficient to jump out of \textit{libc}. In contrast, x86's \texttt{jmp} can reach $\pm$2\,GiB, and PowerPC's \texttt{b} covers $\pm$32\,MiB.

    \item RISC-V instructions are naturally better aligned, so \texttt{nop}s are rarely present in \textit{libc}. This eliminates the possibility of using \textit{nop-trampolines}.

    \item The Linux kernel on RISC-V saves the full context during interrupts, including caller-saved registers. This is relevant for indirect jumps, since a caller-saved register cannot be safely overwritten.
\end{enumerate}

Constraints (1) and (2) necessitate the use of an indirect jump sequence (\texttt{auipc} + \texttt{jalr}). Due to constraint (3), the calling convention must be preserved. An indirect jump requires 8 bytes, plus an additional 8 bytes (if compressed instructions, RVC, are supported) for the prologue and epilogue to store and load the register used for the jump. Creating a 16-byte patch requires four to seven relocatable instructions\footnote{\scriptsize Position-independent instructions that can be safely relocated.}, depending on the level of compression. In contrast, x86 requires 5 bytes when using a long \texttt{jmp}, or only 2 bytes when using a short \texttt{jmp} combined with a \textit{nop-trampoline}. As a result, creating a patch involves replacing only one to three instructions.

\subsection{Solution}
As discussed in subsection~\hyperref[subsec:obstacles]{Obstacles}, each \texttt{ecall} must be surrounded by a sufficient number of relocatable instructions to accommodate a 16-byte patch. This requirement is not always met, for example, some \texttt{ecall}s are surrounded by only a single compressed relocatable instruction, yielding just 6 bytes of usable space. The solution is to treat a nearby \texttt{ecall} with sufficient surrounding space as a gateway, enabling redirection from less spacious locations. Consequently, patches are classified into three types based on the amount of available patch space. The three patching methods are:
\begin{enumerate}
    \item Gateway Patch: Applied when an \texttt{ecall} is surrounded by many relocatable instructions. It creates approximately\footnote{Due to the 2's complement bias, the exact range is from -0x80000800 to 0x7ffff7fe, differing by 4\,KiB.} a $\pm$2\,GiB jump using \texttt{auipc}\,+\,\texttt{jalr}, serving as the foundation for the library. Smaller patches jump to these gateways to reach the \textit{syscall\_intercept} library.
    
    \item Middle Patch: Applied when a sufficient number of relocatable instructions are available to preserve the calling convention. It uses \texttt{jal ra, <gateway address>} to jump to the gateway, where it gets “forwarded” to the \textit{syscall\_intercept} library.
    
    \item Small Patch: Applied when there is not enough space to preserve the calling convention. During the disassembly phase at program runtime, the syscall number (i.e., the immediate value that sets the register \texttt{a7}) is extracted through static analysis and stored in the patch structure. This makes it possible for \texttt{jal} to overwrite \texttt{a7}, since its original value is later restored inside \textit{syscall\_intercept}. The patch uses \texttt{jal a7, <gateway address>} to jump to the gateway, where it gets redirected to the \textit{syscall\_intercept} library.
\end{enumerate}

Patch distribution is approximately 40\%~Gateway, 45\%~Small, and 15\%~Middle, varying by \textit{libc} version.

This system ensures that all syscalls are intercepted by redirecting them to a shared entry point within \textit{syscall\_intercept}, namely the \texttt{asm\_entry\_point} assembly routine. While other architectures jump directly to an allocated space outside \textit{syscall\_intercept}, the RISC-V implementation must dynamically identify each patch based on its unique return address. Once identified, execution is forwarded to the corresponding relocated instruction stored within \textit{syscall\_intercept}.

\subsection{Trade-offs}
Instead of having each patch jump directly to its dedicated location, all patches are redirected to a shared entry point, where they are dynamically identified. This design slightly increases runtime overhead but significantly reduces memory usage and improves cache locality. For more information, check the \hyperref[sec:overhead]{Overhead} section.

\section{Differences between the RISC-V and the x86 version}
The RISC-V version introduces several improvements and differences compared to the x86 version:
\begin{enumerate}
    \item Return Values: The \texttt{syscall\_no\_intercept()} function returns a structure containing values from both the \texttt{a0} and \texttt{a1} registers, the two standard syscall return values. In contrast, the x86 version returns only the primary value (stored in the \texttt{rax} register).

    \item Thread Interception: The RISC-V version intercepts all threads (i.e., all \texttt{clone()} variants) and logs their results, including those with separate stack spaces. The x86 version does not log results for such threads.

    \item Post-clone Hooks: On RISC-V, post-clone hooks are triggered for every thread creation. On x86, these hooks are only triggered for threads with separate stack spaces.

    \item Patching Types: The RISC-V version implements three patch types---Gateway, Middle, and Small---which differ in required patching space and the method used to jump to the \textit{syscall\_intercept} library.

    \item Patching Logic: The patching logic for RISC-V is adapted to the limited range of the \texttt{jal} instruction and the absence of \textit{nop-trampolines}. Additionally, the Linux trap handler on the RISC-V preserves the full register context, requiring explicit save and load operations for the jump register. On x86, the Linux kernel does not preserve caller-saved registers, allowing them to be safely overwritten.
\end{enumerate}

\label{sec:overhead}
\section{Overhead compared to x86}
\subsection{Execution Time Overhead}
To evaluate the execution time overhead of the implementation, three distinct cost scenarios are analysed:
\begin{enumerate}
    \item Normal: The cost of executing the standard \textit{libc} function (e.g., \texttt{getpid()}) without interception. These measurements serve as the baseline reference for overhead calculations.

    \item Intercepted Cost (User Mode): The cost of executing the intercepted \textit{libc} function when the corresponding syscall is bypassed. Since the Linux kernel is not called to execute the syscall, this results in a negative overhead.

    \item Intercepted Cost (Kernel Mode): The cost of executing the intercepted \textit{libc} function with a call to the corresponding syscall. The overhead in this case is positive, due to the complexity of patching \textit{libc} and the added indirection to reach the Linux kernel.
\end{enumerate}

The test results are as follows: On x86, a median overhead of -70\% for (2) compared to (1), and a median of 2\% for (3). On RISC-V, the overhead is -35\% for (2), and 5\% for (3). However, these results are dependent on the execution platform, which in our case includes an i7-8650U x86 CPU (4 cores) from Dell Latitude and a TH1520 RISC-V CPU (4 cores) from Lichee PI 4A.

\subsection{Memory usage}
Regarding memory usage, the RISC-V implementation has a significantly smaller footprint, reduced from 1.37\,MiB to 192\,KiB, representing only 13.6\% of the x86 usage. This figure does not account for the inherently more compact nature of the RISC-V ISA. The reduction comes from the following:
\begin{itemize}
    \item x86 allocates 1\,MiB for templates\footnote{A set of instructions executed before and after entering the C part of \textit{syscall\_intercept} where user-defined \textit{hooks} are.} and relocated instructions, whereas RISC-V requires only 128 KiB for relocated instructions.

    \item x86 allocates 256\,KiB  for the absolute jump trampolines, while RISC-V uses one trampoline with a maximum size of 24\,B.

    \item The bitmap jump table is half the size on RISC-V due to its minimum instruction alignment of two bytes, $\approx$64\,KiB over $\approx$128\,KiB.
\end{itemize}

This reduced memory footprint and improved locality of frequently used code could enhance cache efficiency.

Memory usage on RISC-V scales more favourably. Other architectures allocate memory for a template, relocated instructions, and a dedicated trampoline (i.e., an absolute jump trampoline) for each patch. In contrast, RISC-V leverages a single shared entry point and a shared trampoline. As a result, memory usage scales linearly with the number of patches on RISC-V as \(O(n)\), while x86 scales approximately as \(O(10n)\).

\section{Current status and next steps}
The library is fully functional and is a cost-free replacement for the original library. The next steps include reducing the version gap, particularly by implementing all the missing test cases. A more interesting direction, however, could be to implement a filter that reduces modifications to unused syscalls, tailored to the specific destination use case. For example, in the case of AdHocFS, we could focus on patching only the I/O-related syscalls.

\section*{Acknowledgements}
\tiny{This project is promoted by the Ministry for Digital Transformation and the Civil Service, within the framework of the Recovery, Transformation and Resilience Plan - Funded by the European Union - NextGenerationEU. This work has been partially financed by the European Commission (EU-HORIZON VITAMIN-V GA 101093062). The work carried out in this article was achieved with the support of RISC-V International in their mentorship program, with the participation of Ramon Nou (BSC) as mentor and Petar Andrić as Mentee.}


\printbibliography 


\end{document}